\documentstyle[12pt,epsfig]{article}        
  \advance\oddsidemargin  by -1in
  \advance\evensidemargin by -1in
\newcommand{\Pom}{I$\!\!$P}                
\def\lsim{\mathrel{\rlap{\lower4pt\hbox{\hskip1pt$\sim$}}
    \raise1pt\hbox{$<$}}}         
\def\gsim{\mathrel{\rlap{\lower4pt\hbox{\hskip1pt$\sim$}}
    \raise1pt\hbox{$>$}}}         

\def\Pom{{ I\!\!P  }}
\def\be{\begin{equation}}
\def\ee{\end{equation}}
\def\bq{\begin{eqnarray}}
\def\eq{\end{eqnarray}}

\begin{document}
\pagestyle{empty}

\hfill{\large DFTT xx/96}

\hfill{\large August 1996}

\vspace{2.0cm}

\begin{center}  \begin{Large} \begin{bf}
Longitudinal and Transverse  Nuclear Shadowing\footnote{To be published 
in the Proceedings of the Workshop on ``Future Physics at HERA''.} \\
  \end{bf}  \end{Large}
  \vspace*{1cm}
  \begin{large}
V.~Barone$^a$ and 
M.~Genovese$^{b,}$\footnote{Supported by EU Contract ERBFMBICT
950427.}
\\ 
  \end{large}
\vspace{1cm}

$^a$ Universit{\`a} di Torino and INFN, 10125 Torino, Italy \\
$^b$ Universit\'e Joseph Fourier and IN2P3-CNRS, 38026 Grenoble, 
France \\

\vspace{1cm}

{\large \bf Abstract \\}
\end{center}

Nuclear shadowing arises from multiple scattering of 
the hadronic fluctuations ($\vert q \bar q \rangle, 
\vert q \bar q g ...\rangle$) of the virtual photon in a 
nucleus.
We predict different longitudinal and transverse
shadowing and an $A$--dependence of $R \equiv 
\sigma_L/\sigma_T$ which can be up to a 50\% effect. 
The possibility of detecting nuclear effects on $R$ at HERA 
is discussed. 

\vfill
\pagebreak

\baselineskip 20pt
\pagestyle{plain}

While  
there are very different ideas on the nature of the 
EMC effect at intermediate and large values of $x$, 
the situation seems to 
be less controversial at small $x$, the region of nuclear shadowing 
(NS). There is a general consensus that NS 
is due to the recombination of partons belonging to different 
nucleons. 
A concrete and quantitative realization of this old 
idea has been offered in 
\cite{BGNPZ1} where NS is attributed to the 
multiple scattering  of the hadronic fluctuations of the virtual photon
in nuclei. One can thus speak of parton recombination in that
the multiple scattering diagrams involve quarks and gluons 
which do not belong to a single nucleon. 
This approach leads rather naturally to the prediction
of different shadowing effects in the longitudinal and 
transverse channels and consequently to 
an enhancement of $R \equiv \sigma_L/\sigma_T$ in nuclei \cite{BGNPZ2}.
In the following we shall sketch 
the derivation of this result and present some quantitative estimates. 
We shall also study the possibility of measuring nuclear effects on $R$
at HERA.

%
In virtual-photon--nucleus scattering, using Glauber formalism,
the nuclear cross section is given by 
\be
\sigma^{\gamma^{*}A} = A \, \sigma^{\gamma^{*}N} - 
4 \pi \, \frac{A-1}{A} \, \left. 
\frac{d \sigma^{D}}{dt}\right \vert_{t=0} 
\, \int d^2 \vec b \, \, T^2(\vec b) + \ldots \,,
\label{3}
\ee
where $d \sigma^{D}/dt$ is the $\gamma^{*}N$ diffraction dissociation 
cross section integrated over the mass $M^2$ of the excited hadronic 
states
\be
\left. \frac{d \sigma^{D}}{dt} \right \vert_{t=0} = \int dM^2 \, 
\left. \frac{d^2 \sigma^{D}}{dt dM^2}\right \vert_{t=0} \, 
{\cal F}(k_L^2)\,.
\label{4}
\ee 
The longitudinal 
form factor of the nucleus ${\cal F}(k_L^2)$ appearing in (\ref{4})
suppresses heavy mass excitations corresponding to 
non negligible longitudinal momenta of the recoil proton, 
$k_L = (Q^2 + M^2)/2 \nu = x m_N \, (1 + M^2/Q^2)$. 

Eq.~(\ref{3}) establishes a link between the 
leading nuclear shadowing correction and the pomeron structure function 
which is proportional to $d \sigma^{D}/dt$. This allows relating the 
diffractive DIS currently under experimental study at HERA to the 
small-$x$ nuclear phenomena which will hopefully be a future 
chapter of the HERA program. 

In the Nikolaev--Zakharov picture of small--$x$ DIS \cite{NZ1,GNZ} 
the double
scattering term in (\ref{3}) can be calculated on the basis
of the Fock structure of the virtual photon interacting with
the nucleus. 
At small $x$, the hadronic states 
into which the $\gamma^{*}$ fluctuates ($\vert q \bar q \rangle 
, \vert q \bar q g \ldots \rangle $) have a very long lifetime 
$\sim 1/m_N x$ and their transverse size is frozen during the scattering
process. Hence one can write the virtual 
photoabsorption cross sections for scattering off a nucleon 
$\sigma^{\gamma^{*}N}_{L,T}$ as (focusing for the moment on the 
$q \bar q$ Fock component)
\bq
\sigma^{\gamma^{*}N}_{L,T}(x,Q^2) 
&=& \langle \sigma(\rho, x) \rangle_{L,T} \nonumber \\
&\equiv& \int_0^1 d \alpha  \int d^2 \vec \rho \, \, \vert 
\Psi_{L,T}(Q^2, \rho, \alpha) \vert^2 \, \sigma(\rho, x) 
\,,
\label{5}
\eq
where $\Psi_{L,T}$ are the $q \bar q$ wave functions 
of the virtual photon, $\rho$ the transverse separation of the pair, 
$\alpha$ the momentum fraction carried by one of the  
components, and $\sigma(\rho, x)$ is the interaction
cross section of the $q \bar q$ color dipole with the nucleon, which
does not depend on the flavor and on the photon polarization. 
Glauber's expansion, written in terms of the 
dipole cross section, reads 
\be
\sigma_{L,T}^{\gamma^{*}A}(x,Q^2) = 
A \, \langle \sigma(\rho, x) \rangle_{L,T} -
\frac{A-1}{4 \, A} \, \langle \sigma(\rho, x)^2 \rangle_{L,T}
\, \int d^2 \vec b \, \, T^2(\vec b) + \ldots \,,
\label{6}
\ee
where $\langle \sigma(\rho, x) \rangle_{L,T} \equiv 
\sigma_{L,T}^{\gamma^{*}N}(x,Q^2)$. 
By comparing eqs.~(\ref{3}) and (\ref{6})
one 
identifies the contribution to 
$d \sigma^{D}/dt$ corresponding to the $q \bar q$ content of the 
photon (or of the pomeron, from another viewpoint) as
\be
\left. \frac{d \sigma^{D,q \bar q}_{L,T}}{dt} \right \vert_{t=0} = 
\frac{\langle \sigma(\rho, x)^2 \rangle_{L,T}}{16 \, \pi}\,,
\label{8}
\ee
For small $\rho$, $\sigma(\rho, x)$ has the color transparency
behavior $ \sigma(\rho, x) \propto \rho^2$. 
Because of the structure of $\Psi_{L,T}(Q^2, \rho, \alpha)$ the dominant
contribution to $\sigma_{L,T}^{\gamma^{*}N}$ comes from 
pairs of transverse size 
$\rho^2 \sim [m_q^2 + Q^2 \, \alpha (1-\alpha)]^{-1}$.
Symmetric pairs, with $\alpha \sim 1/2$ and $\rho^2 \sim 1/Q^2$, 
have $\sigma_{L,T} \sim (1/Q^2) \log{(Q^2/m_q^2)}$, {\it i.e.} 
scaling cross sections. 
Asymmetric pairs, with $\alpha \sim 0, 1$ and large size 
$\rho^2 \sim 1/m_q^2$, have a scaling transverse cross section
$\sigma_T \sim 1/Q^2$, but a vanishing longitudinal cross 
section $\sigma_L$. 

Because of the color transparency property of $\sigma(\rho, x)$, 
the second and higher terms in the series (\ref{6}) contain powers
of $\rho^2$. As a consequence, symmetric pairs with 
$\rho^2 \sim 1/Q^2$ give a $1/(Q^2)^2$, {\it i.e.} negligible, 
contribution to the double scattering term, whereas asymmetric pairs
with $\rho^2 \sim 1/m_q^2$ lead to a scaling $1/Q^2$ screening.
Since asymmetric pairs lead to a vanishing $\sigma_L$, 
we can conclude that shadowing in the longitudinal cross section 
is negligible at moderate and large values of $Q^2$, whereas it is 
significant and almost $Q^2$ independent in the 
transverse cross section. 
Thus, $R$ is expected to be enhanced in nuclei for 
$Q^2 \gsim 5$ GeV$^2$. 

So far we have considered only the lowest Fock state 
of $\gamma^{*}$, yielding the component $\propto M^2(M^2 + Q^2)^{-3}$
of the mass spectrum. At large $M^2$ the triple pomeron component,
related to the $q \bar{ q} g$ Fock state of 
the photon, becomes dominant. Its mass spectrum 
is
\be
\left. \frac{d \sigma^{D, 3 \Pom}_{L,T}}{dM^2 \,dt} \right \vert_{t=0} = 
\sigma_{\rm tot}^{\gamma^{*}N} \, A_{3 \Pom} \, 
\frac{M^4}{(Q^2 + M^2)^3}\,.
\label{9}
\ee
Putting all terms together one ends up with 
\be
\sigma_{L,T}^{\gamma^{*}A}(x,Q^2) = 
A \, \sigma_{L,T}^{\gamma^{*}N}(x,Q^2) - 
\frac{A-1}{4 \, A} \, \left \{ \langle \sigma(\rho, x)^2 \rangle_{L,T}
+ 16 \pi \, \left. \frac{d \sigma_{DD}^{3 \Pom}}{dt} \right \vert_{t=0} 
\right \}
\, \int d^2 \vec b \, T^2(\vec b) + \ldots \,,
\label{10}
\ee
A full calculation of the $3 \Pom$ contribution 
has been performed \cite{GNZ}, which has confirmed the behavior (\ref{9}) 
both for the longitudinal and the transverse cross section and 
predicted a coupling $A_{3 \Pom}$ substantially flavor and $Q^2$ 
independent for $Q^2 \gsim 2$ GeV$^2$. 

Our estimate of $\Delta R \equiv R_A - R_N$ in \cite{BGNPZ2}
was based on eqs.~(\ref{10}) and (\ref{9}), with $A_{3 \Pom}$
taken phenomenologically from photoproduction data. Now,
the $A_{3 \Pom}$ value computed in \cite{GNZ} is somehow larger than 
the one used in \cite{BGNPZ2}. This means that, since the triple pomeron 
contribution does not discriminate between longitudinal and 
transverse cross sections, the results
for $\Delta R$ given in \cite{BGNPZ2} slightly overestimate the 
nuclear effects on $R$. We correct them here by an educated guess, 
leaving a precise quantitative determination to a forthcoming 
paper. 

Thus our prediction for $\Delta R \equiv R_A - R_N$ 
in the atomic mass range 
$A \simeq 30 - 80$ (say Cu--Pb), at $x = 10^{-3}$ and $Q^2 = 10$ GeV$^2$ 
is: $\Delta R \simeq 0.10 - 0.15$, that is a $30 - 50 \%$ effect (with
$R_N \simeq 0.30-0.35$). We found that at small $Q^2 \sim 1$ GeV$^2$
shadowing is similar in the longitudinal and transverse cross
sections. 
We expect the largest nuclear effects at $x$ around $10^{-3}$
and $Q^2$ larger than few GeV$^2$.

Let us address the problem of detecting nuclear effects on $R$ at HERA. 
In order to extract $R$ one has to use nucleon beams
with at least two different energies. We consider the possibility
of having two beams with nucleon energies $E_1 = 410$ GeV and 
$E_2 = 205$ GeV. The electron energy is $27.6$ GeV. 

We estimate now the statistical error on 
$\Delta R$. 
By considering two targets $A$ and $B$ and the cross section
ratios $\rho_1 \equiv \sigma_B^{(1)}/\sigma_A^{(1)}$ 
and 
$\rho_2 \equiv \sigma_B^{(2)}/\sigma_A^{(2)}$, 
corresponding to the two target energies, one easily finds the relation
\be
\Delta R  = (\rho - 1) \, (1 + \bar R)\left [ \frac{\rho \, 
(1-z_2)}{1 + z_2 \bar R} - \frac{1-z_1}{1 + z_1 \bar R} \right ]^{-1}
\,,
\label{11}
\ee
among $\Delta R \equiv R_A - R_B$, $\bar R \equiv (R_A + R_B)/2$, 
$z_{1,2} \equiv (1 - y_{1,2})/(1 - y_{1,2} + y_{1,2}^2/2)$, and the 
ratio of cross section ratios $\rho \equiv \rho_1/\rho_2$.   
It is clear from eq.~(\ref{11}) that in order to extract $\Delta R$
one needs $\bar R$, besides 
$\rho$. Since our purpose here is simply to 
evaluate the expected statistical error on $\Delta R$, we
use (\ref{11}) as a constraint between $\bar R$ and $\Delta R$ 
\cite{NMC}. In the following the target $B$ is assumed to be deuteron.
We choose $x =  10^{-3}$,  $Q^2 \simeq 20$ GeV$^2$, 
let $\bar R$ vary in a reasonable range around $0.3-0.4$, 
fix $\rho$
so as to get a $\Delta R$ value around 0.10-0.15 (which 
is our estimate presented above), and 
calculate the error on $\Delta R$. 
The statistical error on $\rho_{1,2}$, 
 with a  
luminosity of 1 pb$^{-1}$ per nucleon,
is taken to be $\delta \rho_{1,2} = 0.0090$ 
 (interpolating the values computed
by Sloan \cite{S}). We set $\rho_1 \simeq 0.80$ and, finally, 
assume a 30\% error on $\bar R$. 
\begin{figure}[tb] \label{FIGUR1}
\mbox{\epsfig{file=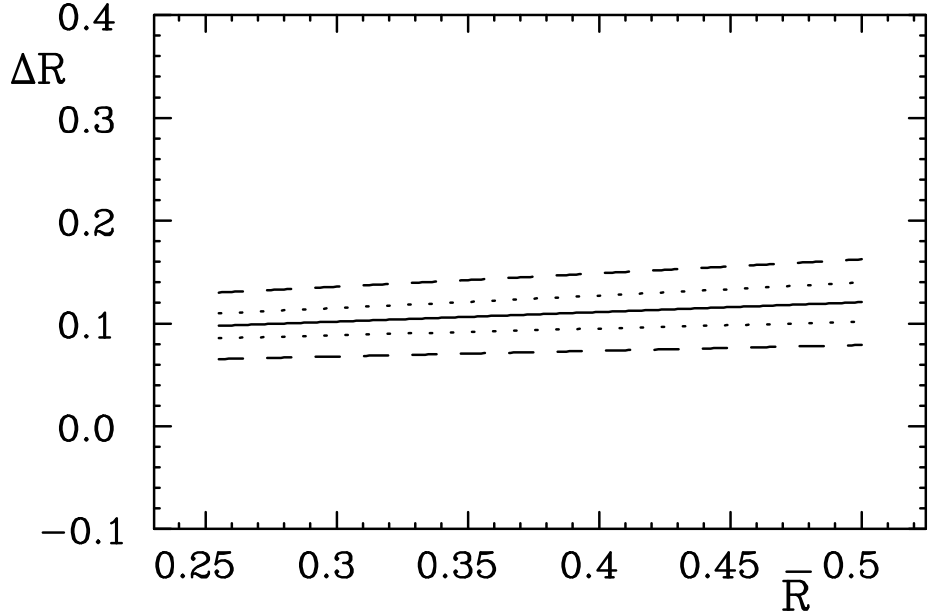,width=0.8\textwidth}}
\caption{$\Delta R$ {\it vs.} $\bar R$ at $x = 10^{-3}$
and $Q^2 = 20$ GeV$^2$.} 
\end{figure}
The result of our evaluation is shown in Fig.~1 where the solid line
represents the central value of $\Delta R$ and 
the dashed and dotted lines mark the 
estimated statistical error with integrated 
luminosities of 1 pb$^{-1}$ and 10 pb$^{-1}$ per nucleon, respectively. 
The estimated statistical uncertainty on $\Delta R$ is thus 
$30-35 \%$. 
Our conclusion is that the nuclear effects on $R$ predicted 
by our model are visible 
at HERA and can be measured with a reasonable accuracy.

Finally we would like to comment on a different approach
to parton recombination \cite{CQR}. 
In the fusion model of \cite{CQR} parton recombination is an 
initial state process and the 
shadowing of 
nuclear structure functions arises from the shadowing 
of the glue density, which is universal, not depending on the 
specific process or observable considered. Thus one would expect
that all gluon--dominated physical quantities, such as 
$F_2^A$ and $F_L^A$ at small $x$, should behave similarly, at variance
with our finding. 
However no quantitative results for $R_A$ and $\Delta R$
have been provided so far for this class of models. It would be
interesting to work out their predictions to see whether 
a possible HERA measurement of $R_A$ and $\Delta R$ 
can also discriminate between different models of nuclear shadowing.

%


\begin{thebibliography}{99}

\bibitem{BGNPZ1}
V.~Barone {\it et al.}, Z. Phys. C58 (1993) 541.

\bibitem{BGNPZ2}
V.~Barone {\it et al.}, Phys. Lett. B304 (1993) 176.




\bibitem{NZ1}
N.N.~Nikolaev and B.G.~Zakharov, Z. Phys. C49 (1991) 607; 
Phys. Lett. B260 (1991) 414.


\bibitem{GNZ}
N.N.~Nikolaev and B.G.~Zakharov, Z. Phys. C64 (1994) 631.
M.~Genovese, N.N.~Nikolaev and B.G.~Zakharov, JETP 81 (1995) 633.



\bibitem{NMC}
P.~Amaudruz {\it et al.} (NMC), Phys. Lett. B294 (1992) 120.

\bibitem{S}
T.~Sloan, The Accuracy of Shadowing Measurements Using Nuclei in HERA, 
contribution to the Workshop on ``Future Physics at HERA''. 

\bibitem{CQR}
F.E.~Close, J.~Qiu and R.G.~Roberts, Phys. Rev. D40 (1989) 2820. \\
S.~Kumano, Phys. Rev. C48 (1993) 2016.







\end{thebibliography}
\end{document}